\documentstyle[psfig]{aa}

\begin{document}

   \thesaurus{11.03.1; 
              11.07.1; 
              13.25.2} 

   \title{Updating the $\sigma$-$T$ relationship for galaxy clusters}

   \author{Xiang-Ping Wu\inst{1},   
           Li-Zhi Fang\inst{2} and Wen Xu\inst{2} 
          }

   \offprints{X. P. Wu}

   \institute{$^1$Beijing Astronomical Observatory, 
        Chinese Academy of Sciences, Beijing 100012, China\\
     $^2$Department of Physics, University of Arizona, Tucson, AZ 85721, USA
             }

   \date{Received 15 October, 1997; accepted 13 August, 1998}

   \maketitle

   \begin{abstract}\footnote{Table 1 is only availbale in electronic 
                form at the
                CDS via anonymous ftp to cdsarc.u-strasbg.fr 
                (130.79.128.5) or via 
                http://cdsweb.u-strasbg.fr/Abstract.html}

        The relationship between the X-ray determined temperature $T$
        of the intracluster gas 
        and the optical measured velocity dispersion $\sigma$ of
        the cluster galaxies
        is often believed to be not only a straightforward but also 
        robust test for the dynamical properties of galaxy clusters. 
        Here, we present the $\sigma$-$T$ relationship using 
        the 94 clusters drawn from the largest sample of 149 clusters 
        in literature, for which both $\sigma$ and $T$ 
        are observationally determined. 
        Employment of the doubly weighted orthogonal distance regression 
        to our sample yields
        $\sigma=10^{2.47\pm0.06}T^{0.67\pm0.09}$, indicating an
        apparent deviation of dynamical state from that predicted 
        by the isothermal and hydrostatic 
        equilibrium model for galaxy clusters,
        though the average ratio $\beta_{spec}$ of 
        specific energy in galaxies to that in gas is found to be
        in excellent agreement with unity.
        It shows that a nonisothermal gas distribution 
        with a mean polytropic index of $\gamma=1.3$ can account 
        for the reported  $\sigma$-$T$ relationship,  
        while overall clusters 
        can still be regarded as dynamically-relaxed systems.


   \end{abstract}

%

\section{Introduction}

Clusters of galaxies are the largest coherent and gravitationally
bound objects in the universe. They are often used for cosmological
test of theories of formation and evolution of structures
in the universe (e.g. Bahcall \& Cen 1993). 
They also play a potentially important role in 
the direct measurement of the present mean mass density of 
the universe (e.g. Bahcall, Lubin \& Dorman 1995).
Yet, these cosmological applications are closely connected with
the question of how reliable our current knowledge about the dynamical 
properties of clusters would be.  Numerous recent observations made
primarily at X-ray waveband suggest that the gross dynamical properties 
of galaxy clusters have experienced little evolution since redshift
$z\sim0.8$ (e.g. Mushotzky \& Scharf 1997; Rosati et al. 1998; 
Vikhlinin et al. 1998). A statistical comparison of 
different cluster mass estimators from optical/X-ray and gravitational
lensing measurements  shows that, regardless of the presence of
substructures and local dynamical activities, 
overall clusters of galaxies
even at intermediate redshift can be regarded as 
dynamically-relaxed systems (e.g. Allen 1998; Wu et al 1998
and references therein).  This essentially justifies the employment of
hydrostatic equilibrium for galaxy clusters. 

One simpler and more straightforward approach to testing the
dynamical state of clusters of galaxies, which was suggested two decades
ago (Cavaliere \& Fusco-Femiano 1976), 
is to use the relationship  between the X-ray determined temperature $T$
of the intracluster gas and the optical measured velocity dispersion 
$\sigma$ of the cluster galaxies:  If both galaxies and gas
are the tracers of the depth and shape of a common gravitational potential, 
we would expect $\sigma\sim T^{0.5}$. In other words, the disagreement 
between the observed $\sigma$-$T$ relationship and 
the expectation of $\sigma\sim T^{0.5}$ may be considered to be a strong 
indicator for the departure of cluster dynamical state 
from the isothermal hydrostatic equilibrium. 
With the rapid growth of optical/X-ray data for clusters of galaxies
over the past years, a number of authors (see Table 2) have made attempt 
at determining the $\sigma$-$T$ relationship from various cluster samples.
While the resultant $\sigma$-$T$ profiles
are not inconsistent with  $\sigma\sim T^{0.5}$, 
one cannot exclude the possibility that there is a real 
deviation of the observed $\sigma$-$T$ relationship from 
that expected under the scenario of  isothermal hydrostatic equilibrium.  
For instance, based on the well-defined
cluster samples in which the measurement errors are well determined, 
Bird, Mushotzky \& Metzler (1995), Girardi et al (1996) and
White, Jones \& Forman (1997) found 
$\sigma\sim T^{0.6}$. Yet, a difficulty with these determinations
arises from the too small cluster samples, which often have 
large data scatters in the fit of  $\sigma$-$T$ relationship.
It appears that a larger cluster sample is needed  
in order to achieve a better statistical significance, with which 
one can determine whether there is an intrinsic dispersion 
in the $\sigma$-$T$ relationship due to different physical mechanisms
among different clusters. We wish to fulfill the task in the 
present paper by updating the $\sigma$-$T$ relationship for clusters,
making use of all the published data sets of $\sigma$ and $T$ in literature.
We examine the question whether the $\sigma$-$T$ relationship can be used 
for the purpose of testing the dynamical properties of clusters.

\section{Sample}

By extensively searching literature, we find 517 galaxy clusters for
which the velocity dispersions (367 clusters)  and/or temperatures (299
clusters) are observationally determined.  Here we exclude those clusters
whose $\sigma$ or $T$ are obtained by indirect methods such as  the
$L_x$-$\sigma$ and $L_x$-$T$ correlations (e.g. Ebeling et al. 1996; 
White et al. 1997) where $L_x$ is the X-ray luminosity,
or by the gravitational lensing analysis (e.g. Sadat, Blanchard \&
Oukbir  1998). 
The final sample in which both $T$ and $\sigma$ are
available for each cluster contains 149 clusters (Table 1). 
This compares with
the similar but the largest cluster sample (207 clusters) heretofore 
published by White et al (1997), in which 83 clusters have the measured 
$\sigma$ and $T$, thus we have increased the data set by  66 clusters.
Since the data of $\sigma$ and $T$ are collected among a large number of
individual sources, it may be too tedious to list all the references
in the present paper. For the majority of the data the  reader is referred to
Zabludoff, Huchra \& Geller (1990), 
Struble \& Rood (1991),
Edge \& Stewart (1991), Lubin \& Bahcall (1993), Gioia \& Luppino (1994), 
White \& Fabian (1995), Ebeling et al.(1996), 
Carlberg et al. (1996), Fadda et al. (1996),  
Girardi et al. (1996), Girardi et al. (1997), 
Smail et al. (1997), Mushotzky \& Scharf (1997), 
Wu \& Fang (1997),  Ettori, Fabian \& White (1997) 
and White et al. (1997). 

   \begin{table}
      \vspace{1cm}
      \caption{Cluster sample (see CDS)}
         \label{Table 1}
   \end{table}

\section{The $\sigma$-$T$ relationship}

Fig.1 shows velocity dispersion $\sigma$ versus temperature $T$
for the 149 clusters listed in Table 1, and it is clearly seen that 
there is a strong correlation between the two variables. 
Since not all the measurement uncertainties in $\sigma$ and $T$ are known 
in our cluster sample, we first employ the standard ordinary least-square 
(OLS) fit of a power-law to all the data set, which yields
\begin{equation}
\sigma=10^{2.54\pm0.03}T^{0.56\pm0.04},
\end{equation}
where (also hereafter) $\sigma$ and $T$ are in units of km s$^{-1}$ and
keV, respectively. Note that the error bars in eq.(1)  do not include
the measurement uncertainties. 
We then use the orthogonal distance regression (ODR)
technique ODRPACK (Bogg et al. 1989; Bogg \& Rogers 1990;
Feigelson \& Babu 1992) to fit a subsample of 94
clusters, for which the measurement
uncertainties in both variables are observationally given (Fig.2). 
The doubly weighted ODR $\sigma$-$T$ relationship reads
\begin{equation}
\sigma=10^{2.47\pm0.06}T^{0.67\pm0.09}.
\end{equation}
Here the quoted  $1\sigma$ standard deviations are determined 
by the Monte-Carlo simulation which has taken the measurement
uncertainties in both $\sigma$ and $T$ into account. 
It turns out that
at about $95\%$ confidence interval, we have detected a 
deviation of the $\sigma$-$T$ relationship from that
expected in the framework of isothermal and hydrostatic equilibrium
for clusters of galaxies, although the parameter scatters are only
slightly reduced as compared with the recent result of 
White et al. (1997).

Nevertheless, as it has been noticed before (Bird et al. 1995),  
the resultant $\sigma$-$T$ relationships seem to depend also on the 
adopted linear regression methods: The OLS method often provides a relatively 
smaller power index in the fit than other regression methods 
such as the Bisector and ODR,  which is independent of 
whether the data are weighted or not. For instance, 
employing the unweighted ODR method to the whole data set of 149 clusters
without including the measurement uncertainties, we can still reach a 
power-law of large index:    
\begin{equation}
\sigma=10^{2.47\pm0.03}T^{0.67\pm0.04}.
\end{equation}
in contrast with the result of eq.(1) obtained by OLS.
Indeed, one may arrive at very 
different conclusions regarding the dynamical properties of clusters 
based on the different fitting methods.
This accounts for the early claim by
Lubin \& Bahcall (1993), in terms of their fitted relationship 
$\sigma\propto T^{0.5}$ by OLS,  
that the overall clusters can be regarded as well virialized
and isothermal systems. So, we may need  to have a close examination of
the working hypotheses in the two fitting methods. 
Isobe et al. (1990) listed four assumptions under which OLS 
method may hold. One of them requires that the values of the 
independent variable (i.e. $T$ in our problem) are measured 
without error. Namely, in the linear fit of the $\sigma$-$T$ relationship,
OLS ignores the scatters around $T$ and only minimizes the residuals 
in $\sigma$.  On the other hand, ODR method is advocated to deal with  
the following question (Feigelson \& Babu 1992):  
What is the intrinsic relationship between properties
$X$ and $Y$ in these objects, without treating one variable differently
from the other ? In other words, ODR makes an attempt
at accounting for data scatters around both $T$ and $\sigma$ 
for our particular problem. Therefore, it appears that in principle
 OLS cannot be used in the fitting
of the $\sigma$--$T$ relationship for clusters, in which both $\sigma$
and $T$ contain significant measurement uncertainties in addition to
their intrinsic scatters.

   \begin{table}
      \vspace{5cm}
      \caption{Summary of the best fitted $\sigma$-$T$ relationships}
          \label{Table 2}
   \end{table}

    \begin{figure} 
     \vspace{5cm}
      \caption{The $\sigma$-$T$ relationship for the 149 clusters in
         Table 1. The low-redshift ($z<0.1$) and high-redshift
         ($z\geq0.1$) clusters are represented by the open triangles (110) and 
         the filled squares (39), respectively. The solid line is the best 
         OLS fitted relationship to the data.}
         \label{fig.1}
   \end{figure}
    \begin{figure}
      \vspace{5cm}
      \caption{The $\sigma$-$T$ relationship of the 94 clusters for which
         the measurement uncertainties in both $\sigma$ and $T$ are known.
         The solid line shows the best OLS fit to the data, and 
         the dashed line is the doubly weighted ODR result.}
       \label{fig.2}
   \end{figure}

We now examine whether the  $\sigma$-$T$ relationship evolves with cosmic
epoch. To this end, we divide our cluster sample of 149 clusters
into two subsamples according
to redshifts:  110 clusters at low redshift $z<0.1$ and  39 clusters 
at high redshift $z\geq0.1$. The best OLS fitted $\sigma$-$T$ relationships
for these two subsamples are
\begin{equation}
 \begin{array}{ll}
\sigma=10^{2.57\pm0.03}T^{0.49\pm0.05}; & \;\;\;\;\;z<0.1;\\
\sigma=10^{2.57\pm0.08}T^{0.56\pm0.09}; & \;\;\;\;\;z\geq0.1.
 \end{array}
\end{equation}
So, within $1\sigma$ error bars there is no apparent change
in the $\sigma$-$T$ relationship between the low-redshift subsample
and the high-redshift one. This is consistent with the recent result of
Mushotzky \& Scharf (1997) based on 13 high-redshift clusters at $z>0.14$ 
whose $\sigma$ and $T$ are well measured. Contrary to the fitting of
the $\sigma$-$T$ relationship, the study of whether 
the $\sigma$-$T$ relationship varies with redshift is insensitive to
the adopted linear regression method as long as we use the same fitting
technique. Yet, the actual reason why we did not apply the ODR method to
our subsample to do the above exercise is the sparse data of the 
high-redshift clusters.  Nevertheless, a visual examination of Fig.2 
reveals that the distributions of the high and low redshift clusters
over the $\sigma$-$T$ plot do not exhibit significant differences.

\section{The $\beta$ parameter}

An equivalent parameter that has been frequently used to 
characterize the dynamical properties of clusters is 
the ratio of specific kinetic energy in galaxies to that in gas:
\begin{equation}
\beta_{spec}\equiv\frac{\sigma^2}{kT/\mu m_p}, 
\end{equation}
where $\mu m_p\approx0.59$ is the average particle mass. 
In the framework of the standard isothermal 
hydrostatic model for clusters, it is expected that
$\beta_{spec}=1$. Fig.3 shows the resulting $\beta_{spec}$ 
from the observed $\sigma$ and $T$ of each cluster lasted in Table 1, 
and the mean value is $\langle \beta_{spec}\rangle=1.00\pm0.52$, 
indicative of a perfect energy equipartition between the galaxies and gas 
in clusters.  This value is also consistent with the result found
from the recent N-body/gasdynamical simulations of formation and evolution
of X-ray clusters (Eke, Navarro \& Frenk 1998).  Because there is an 
apparent asymmetry in the distribution of the $\beta_{spec}$ values in 
Fig.3, we also provide the median and $90\%$ limits of the distribution:
$\beta_{spec}=0.80^{+1.04}_{-0.52}$. 

    \begin{figure} 
      \vspace{5cm}
      \caption{The  parameter $\beta_{spec}=\sigma^2/(kT/\mu m_p)$  
        for the 149 clusters in Table 1. The dotted line
        shows the mean value $\langle \beta_{spec}\rangle$=1.00.}
      \label{fig.3}
   \end{figure}

As an analogy to eq.(4),  Fig.4 gives another way to demonstrate if  
clusters have undergone a significant evolution: the mean value
$\beta_{spec}$ versus redshift. It appears that the whole data set 
is consistent with a nonevolutionary scenario, though there is  
a scarcity of high-redshift clusters in the present sample. 
Our doubly weighted ODR relationship reads 
\begin{equation}
\beta_{spec}=10^{0.00\pm0.09}(1+z)^{0.10\pm0.86}.
\end{equation}

    \begin{figure}
      \vspace{5cm}
      \caption{The binned value $\beta_{spec}$ is plotted against
        redshift. Each redshift bin except the last point (9 clusters)
        at large redshift contains 10 clusters.}
       \label{fig.4}
   \end{figure}

\section{Discussion}

There are many mechanisms that can lead the $\sigma$-$T$ relationship
of galaxy clusters to deviate from that predicted by the 
isothermal and hydrostatic equilibrium model.
These include the anisotropy of galaxy velocity dispersion $\sigma$,
the protogalactic winds which heat the intracluster medium, the velocity
bias between galaxies and dark matter particles, the asymmetric mass
distributions (Bird et al. 1995; Girardi et al. 1996) and 
the effect of cooling flows (White et al. 1997).   
Here, we explore another possibility: The intracluster gas is nonisothermal
and the gas temperature declines with outward radius.  This is 
primarily motivated by the recent spatially resolved X-ray spectroscopic 
measurements of many clusters, which show a significant
radial temperature decline at large radii (e.g. Briel \& Henry 1994;
Henriksen \& White III 1996; Markevitch et al. 1997; etc).
We drop the isothermal assumption but maintain the hydrostatic equilibrium 
model for clusters. The last point is justified by a number of numerical and 
observational studies which indicate that on average, 
galaxy clusters are regular objects with little cosmological and dynamical 
evolution within $z\approx1$. The following exercise is to 
demonstrate how a similar relationship to the 
observationally fitted one in the present paper, $\sigma\propto T^{0.67}$, 
is obtained in the framework of nonisothermal gas distribution.

We use a polytropic temperature profile 
$T=T_0[n(r)/n_0]^{\gamma-1}$
and a conventional $\beta$ model with $\beta_{fit}=2/3$ 
for the gas distribution
$n(r)=n_0[1+(r/r_c)^2]^{-3\beta_{fit}/2}$, where $n_0$ and $r_c$ are
the central gas number density and the core radius, respectively.
Strictly speaking, a $\beta$ model of $\beta_{fit}=2/3$ corresponds to
a polytropic index of $\gamma=1$ (Cowie, Henriksen \& Mushotzky 1987).
Our choice of the density profile with  $\beta_{fit}=2/3$ is only to 
parameterize the observed characteristics. 
If the X-ray emission of a cluster is detected over
a circular aperture of radius $r_{cut}$  from the cluster
center due to the limitation of instruments,  
we would expect an emission-weighted temperature $T_{weighted}$
instead of the temperature profile $T(r)$ when the spatially resolved
spectroscopy is not available:
\begin{equation}
T_{weighted}=\frac{\int_0^{r_{cut}}\alpha(T)Tn^2(r)r^2dr}
                  {\int_0^{r_{cut}}\alpha(T)n^2(r)r^2dr},
\end{equation}
where $\alpha(T)$ is cooling function  and can be approximately
taken to be $\alpha\propto T^{1/2}$ for the thermal bremsstrahlung
radiation. $r_{cut}$ corresponds to the edge of 
the X-ray surface brightness $S(r)$ that the detector can reach:
\begin{equation}
S(r_{cut})\propto\int_{r_{cut}}^{\infty}\alpha(T)n^2(r)
                \frac{rdr}{\sqrt{r^2-r_{cut}^2}}.
\end{equation}
Similarly, the observed velocity dispersion of cluster galaxies 
is the so-called average line-of-sight velocity dispersion weighted
by the galaxy surface number density profile over a certain aperture. 
However, unlike the X-ray emitting gas, the spatial distribution of 
galaxies is more concentrated toward the cluster center and 
their radial velocity dispersions exhibit no significant decline 
at the corresponding radius (e.g. Carlberg, Yee \& Ellingson 1997). 
For simplicity, the velocity dispersion of galaxies is assumed 
to be constant.

We estimate the gas temperature $T=T_c$
at $r=1.25r_c$ using $T_c=(\mu m_p)\sigma^2$ so that the amplitude
of $T(r)$ can be fixed when the polytropic index $\gamma$ 
is specified.  This choice is somewhat arbitrary. 
Our main consideration is to ensure that 
the theoretically expected emission-weighted temperature $T_{weighted}$
is essentially consistent with the observationally
determined one $T_{obs}$. Namely, the position of $r=1.25r_c$ 
is determined by requiring  
$\langle T_{weighted}\rangle=\langle T_{obs}\rangle$ on average
for the 94 clusters in Fig.2 (see also Fig.5), in which  
we take a mean value of the polytropic index, $\gamma=1.3$,
for all the clusters according to the recent measurements of 30 nearby
clusters by Markevitch et al. (1998) as well as some theoretical
considerations (Chiueh \& Wu 1998).  If clusters are 
selected above the same threshold $S(r_{cut})$, $r_{cut}$ can be 
found for each cluster of different temperature $T(r)$, 
where an absolute calibration is needed in order to test
the correlation between the velocity dispersion $\sigma$ and 
the emission-weighted temperature $T_{weighted}$ within $r=r_{cut}$. 
We adopt such a calibration that for a central temperature 
$T_0=5$ keV cluster, $r_{cut}=2r_c$. We utilize the data set of 
velocity dispersion in the subsample of 94 clusters,
in which the measurement uncertainties are known, to estimate 
the emission-weighted temperature $T_{weighted}$ for
each cluster in terms of eqs.(7) and (8), and the resultant 
$T_{weighted}$ versus $T_{obs}$ is displayed in Fig.5. 
Now, it is straightforward to plot the emission-weighted temperature
$T_{weighted}$ against the velocity dispersion $\sigma$ for each of the 94
clusters (Fig.6).  In our model a higher temperature cluster often  
exhibits a larger spatial extension in X-ray emission, in which the
emission-weighted temperature over the whole cluster is smaller than 
$T_c$, while with the same detection threshold,
the overall cluster temperature would appear to be  higher than $T_c$   
if the cluster has a X-ray emitting region of radius $r<1.25r_c$. 
This is the basic scenario illustrated in Fig.6. 
Employing the doubly weighted ODR method to the
data set gives
\begin{equation}
\sigma=10^{2.47\pm0.07}T^{0.67\pm0.07},
\end{equation}
Meanwhile, the average value $\beta_{spec}$ according to
$\sigma$ and $T_{weighted}$ is $\beta_{spec}=0.99\pm0.18$.
These results are in good agreement with those found from the observed data.
Of course, the real
situation is much more complex. For examples,
the absolute calibration of $r_{cut}=2r_c$ for $T_0=5$ keV does not
hold true for all the clusters selected with very different methods,
and the kink at $T=3.6$ keV in the simulated data has actually 
arisen from this poor calibration.
Moreover, the flux threshold $S(r_{cut})$ does not remain to be
constant in different observations, and it can be also 
affected by background emission. 
Although our analysis is only illustrative, it indeed
shows that the nonisothermal 
intracluster gas can be one of the major reasons for the departure of 
the $\sigma$-$T$ relationship from that expected under the 
scenario of isothermal hydrostatic equilibrium.

    \begin{figure} 
      \vspace{5cm}
      \caption{Comparison of the theoretically estimated gas 
        temperature $T_{weighted}$ in terms of eqs.(7) and (8)
        with the observationally determined value  
        $T_{obs}$ for a subsample of 94 clusters shown in Fig.2.
        A $\beta$ density profile 
        with $\beta_{fit}=2/3$ and a polytropic temperature distribution
        with $\gamma=1.3$ are used.  Absolute calibrations are made
        through $T(1.25r_c)=(\mu m_p)\sigma^2$ and $T_0=T(0)=5$ keV for 
        $r_{cut}=2r_c$. All the clusters are assumed to be detected
        above the same threshold of brightness. The dotted line is 
        obtained assuming that $T_{weighted}=T_{obs}$. }
        \label{fig.5}
   \end{figure}

    \begin{figure}
      \vspace{5cm}
      \caption{The correlation between the emission-weighted 
        temperature $T_{weighted}$  and the velocity dispersion $\sigma$
        for the 94 clusters in Fig. 2 (triangles). 
        The solid line represents the best-fitted ODR $\sigma$-$T$ 
        relationship, $\sigma=10^{2.47\pm0.07}T^{0.67\pm0.07}_{weighted}$. 
        We also illustrate  the isothermal case (squares) in which 
        $T=T_c=(\mu m_p)\sigma^2$ (dotted line). }
         \label{fig.6}
   \end{figure}

\section{Conclusion}

We have updated the $\sigma$-$T$ relationship for clusters of galaxies
using 149 published data sets of velocity dispersions and temperatures  
in literature, among which a subsample of 94 clusters is formed 
where  measurement uncertainties in the two variables are 
known. These constitute the largest cluster samples ever used 
for such a kind of analysis.  
The previous claim that the fitted  $\sigma$-$T$ relationship is  
consistent with an isothermal hydrostatic scenario is found to 
have arisen from an inappropriate application of 
the OLS linear regression method.
A plausible fit of the  $\sigma$-$T$ relationship has been achieved 
by using the doubly weighted ODR technique for a subsample of 94 clusters,
which gives rise to $\sigma\propto T^{0.67\pm0.09}$. 
At a high significance level of $\sim95\%$ 
we have detected a deviation of the $\sigma$-$T$ relationship
from that predicted by the isothermal and hydrostatic equilibrium
model, though the mean value of the ratio $\beta_{spec}$ between 
the specific energy of the galaxies and the gas 
is in excellent agreement with unity.
We suggest that such a deviation may be due to the
gas temperature decline at large cluster radius. This has been justified
by a simple theoretical analysis, in which the gas temperature drop
at large radius is described by a polytropic index of $\gamma=1.3$.
Yet, the real situation may be more complex than 
the nonisothermal temperature distribution  for intracluster gas.
A number of factors can also contribute to the reported deviation of 
$\sigma$-$T$ relationship from $\sigma\propto T^{0.5}$ 
(Bird et al. 1995; Girardi et al. 1996; White et al. 1997)). 
Further work will thus be needed to explore the mechanism of
this departure.

\begin{acknowledgements}
We thank the referee, David A. White, for many valuable suggestions
and comments. 
This work was supported by the National Science Foundation of China, 
under Grant No. 19725311.
Wen Xu acknowledges a World Laboratory fellowship.

\end{acknowledgements}

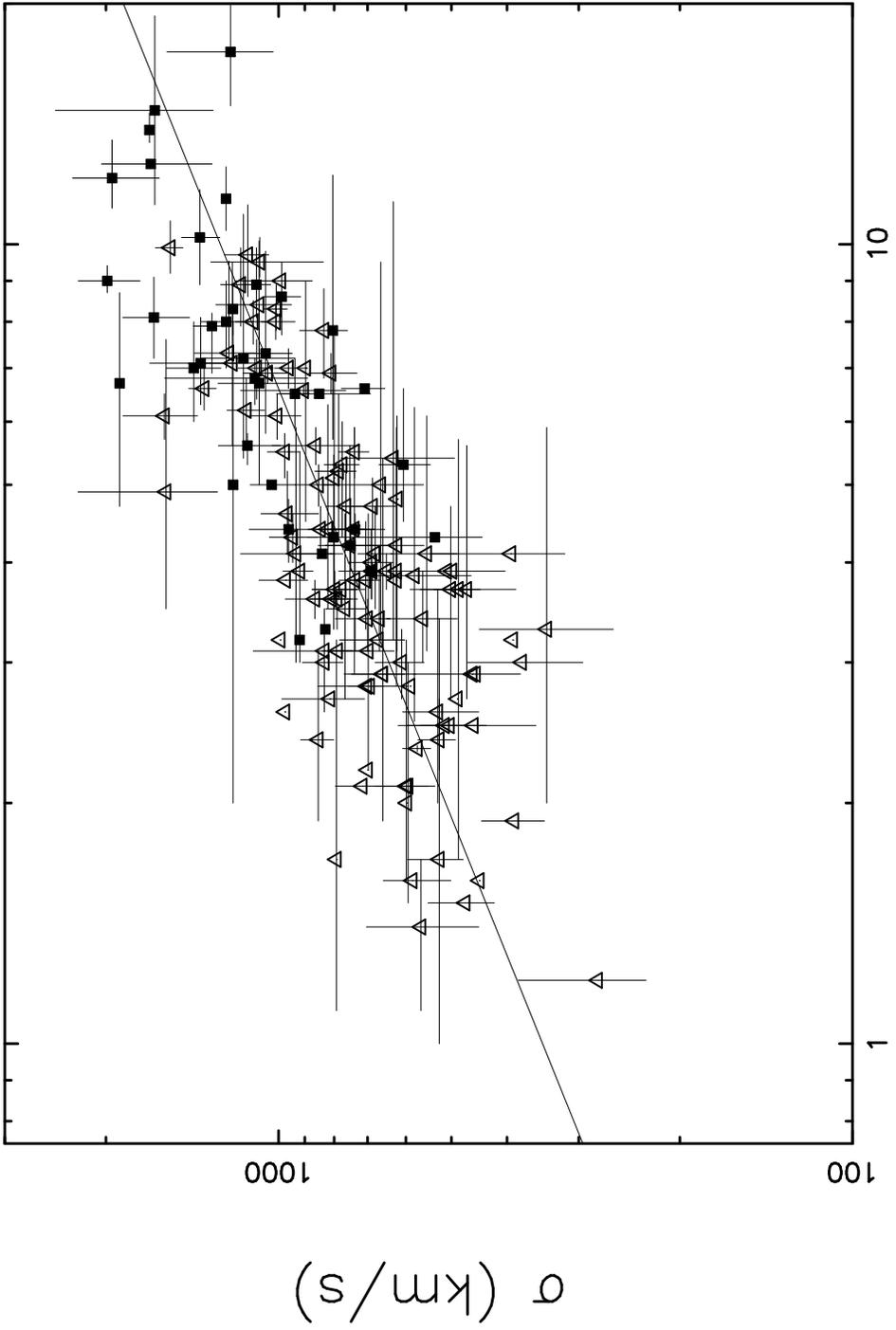
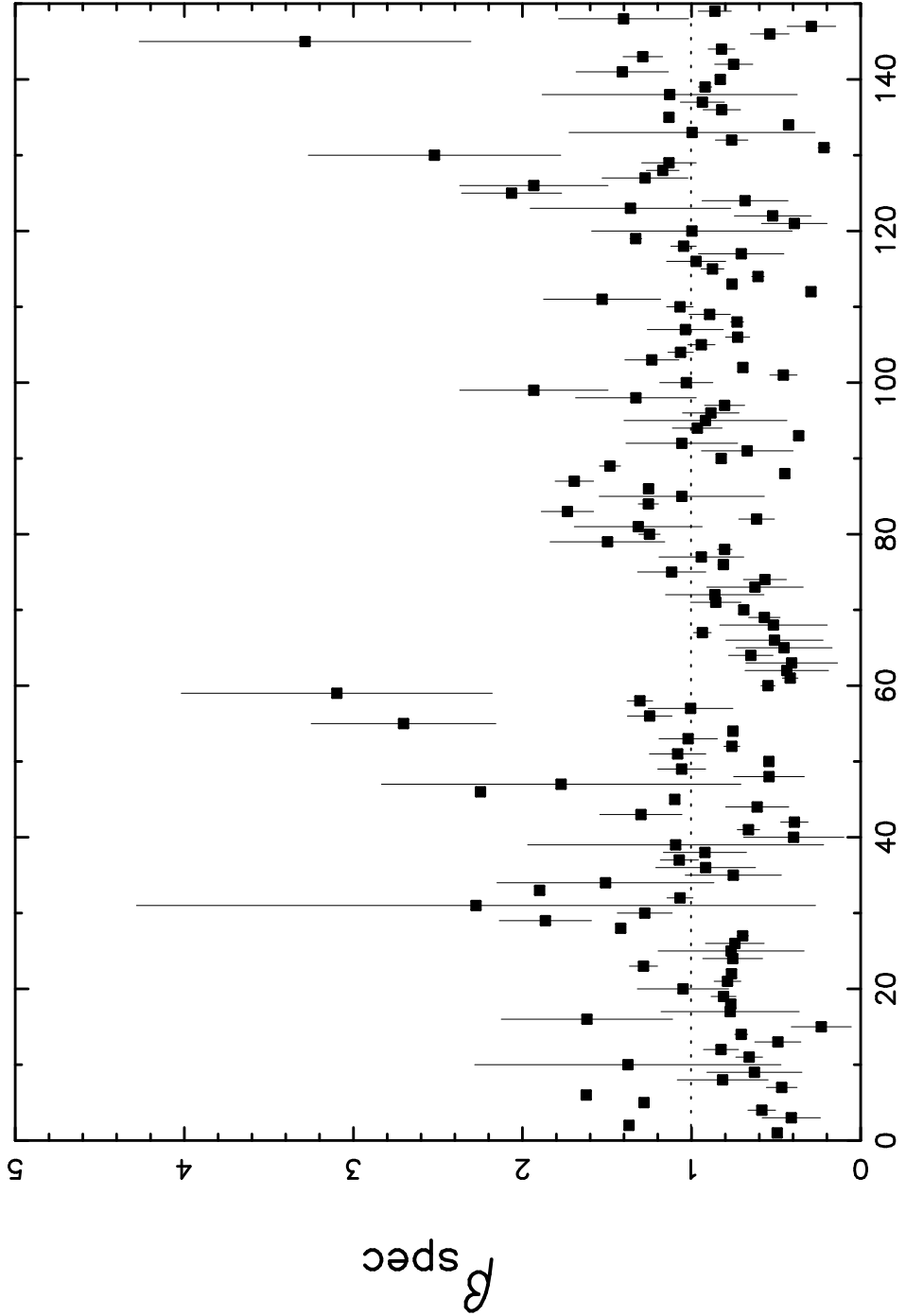
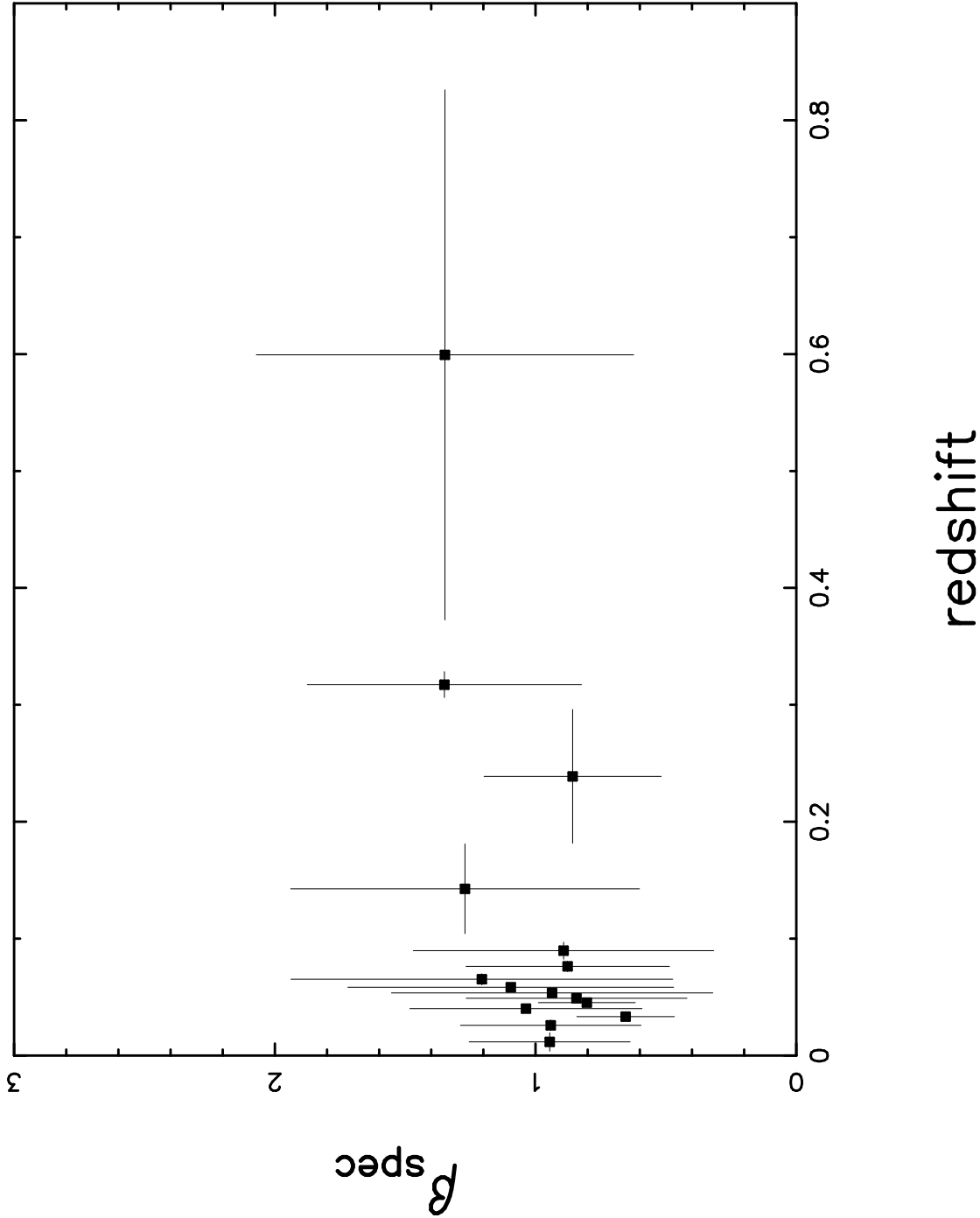
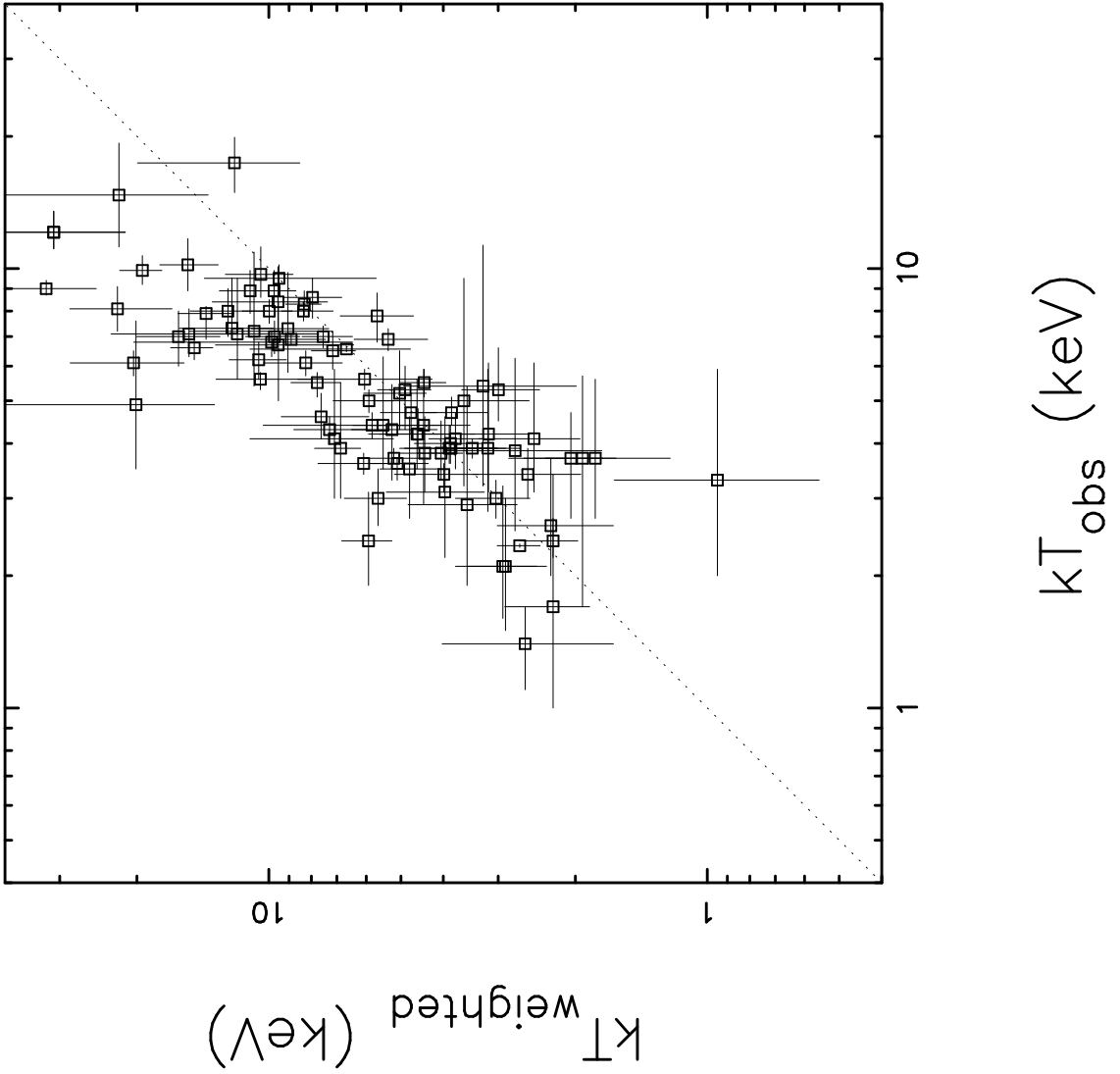
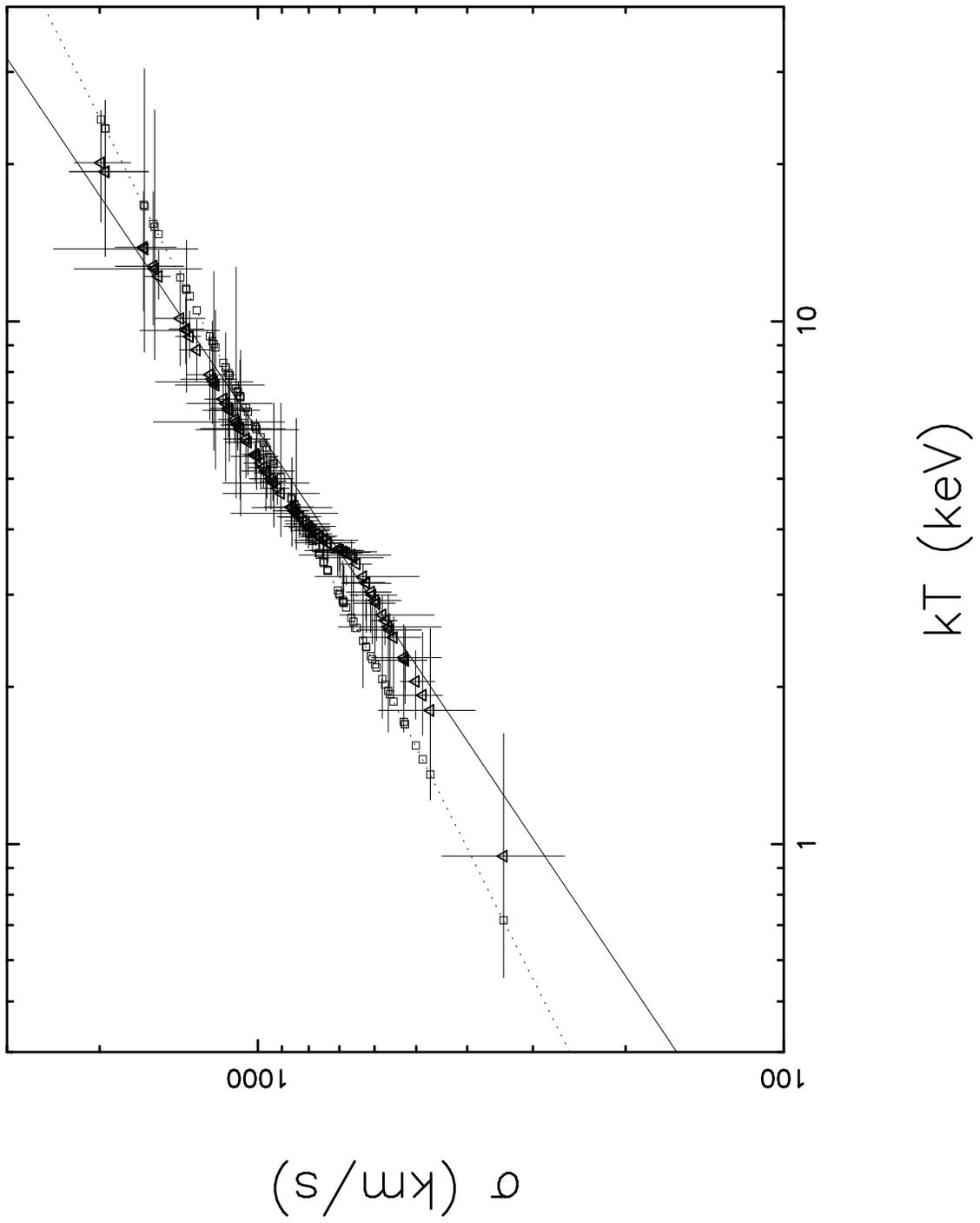
\end{document}